\documentstyle[preprint,aps,version2]{revtex}
\begin{document}
\newcommand{\Tr}{\mbox{Tr}}
\newcommand{\slh}{\hspace{-.5em}/}
\newcommand{\qq}{Q^2}
\newcommand{\oas}{\mbox{$\mbox{O}(\alpha_{s})$}}
\newcommand{\oasz}{\mbox{$\mbox{O}(\alpha_{s}^{2})$}}
\newcommand{\as}{\mbox{$\alpha_{s}$}}
\newcommand{\asz}{\mbox{$\alpha_{s}^{2}$}}
\newcommand{\lms}{\mbox{$\Lambda_{\overline{\mbox{\tiny MS}}}$}}
\def\Lms#1{\Lambda_{\overline{MS}}^{(#1)}}
\def\asp{{\alpha_s\over\pi}}
\def\ams{\alpha_{\overline {MS}} }
\font\fortssbx=cmssbx10 scaled \magstep2
\hbox to \hsize{
\setlength{\footskip}{1.5cm}
\frenchspacing
\everymath={\displaystyle}
  \def\thebibliography#1{{\bf{References}}\list
 {[\arabic{enumi}]}{\settowidth\labelwidth{[#1]}\leftmargin\labelwidth
   \advance\leftmargin\labelsep
   \usecounter{enumi}}
   \def\newblock{\hskip .11em plus .33em minus -.07em}
   \sloppy
   \sfcode`\.=1000\relax}
  \let\endthebibliography=\endlist
%\special{psfile=/NextLibrary/TeX/tex/inputs/uwlogo.ps
%			      hscale=8000 vscale=8000
%			       hoffset=-12 voffset=-2}
\hskip.5in \raise.1in\hbox{\fortssbx University of Wisconsin - Madison}
\hfill$\vcenter{\hbox{\bf MAD/PH/834}
                \hbox{\bf UCD--94--23}
                \hbox{June 1994}}$ }
\vspace{.5in}

\begin{title}
$W$ and $Z$ Polarization Effects in Hadronic Collisions
\end{title}

\author{E.~Mirkes$^a$ and J.~Ohnemus$^b$}

\begin{instit}
$^a$Physics Department, University of Wisconsin, Madison, WI 53706, USA\\
$^b$Physics Department, University of California, Davis, CA 95616, USA
\end{instit}

\begin{abstract}
A Monte Carlo study of the polar and azimuthal angular distributions
of the lepton pair arising from the decay of a $W$ or $Z$ boson
produced at high transverse  momentum in hadronic collisions is
presented. In the absence of cuts on the final state leptons, the
lepton angular distribution in the gauge boson rest frame is
determined by the gauge boson polarization.  Numerical results for the
lepton angular distributions in  the Collins--Soper frame with
acceptance cuts and energy resolution smearing applied to the leptons
are presented.  In the presence of cuts, the lepton angular
distributions are dominated by kinematic effects rather than
polarization effects, however, some polarization effects are still
observable on top of the kinematic effects. Polarization effects are
highlighted when the experimental distributions are divided by the
Monte Carlo distributions obtained using isotropic gauge boson decay.
\end{abstract}
\thispagestyle{empty}
\newpage

\section{INTRODUCTION}

The measurement of the angular distribution of leptons  from the decay
of a gauge boson $V\rightarrow l_1 l_2$ [$V = W, Z$] produced in
hadronic collisions via  a Drell-Yan type process $h_1+h_2\rightarrow
V+X$ provides a detailed test of the production mechanism of the gauge
boson.  In the absence of cuts on the final state leptons, the  lepton
angular distribution in the gauge boson rest frame is determined by
the gauge boson polarization. At ${\cal O}(\alpha_s)$ in perturbative
QCD the angular distribution is described by six helicity cross
sections, which are functions of the transverse momentum and rapidity
of the gauge boson.  The lepton angular distribution can be used to
analyze the gauge boson polarization and thereby test the underlying
production dynamics in much more detail than is possible by rate
measurements alone.

The hadronic production of a vector boson $V$ is described in  lowest
order [$O(\alpha_s^0)$]  by the Drell-Yan quark-antiquark annihilation
process $q\bar{q}\rightarrow V$. In the case of $W$ boson production
at low transverse momentum $p_T^{}$, the  $W$ bosons are produced
fully polarized along the beam direction due to the pure $V-A$
coupling of the charged currents, and the resulting angular
distribution of the charged lepton in the $W$ boson rest frame is
simply
\begin{equation}
dN/d(\cos\theta) \sim
 \mbox{valence quarks}\otimes(1+\cos\theta)^2
+\mbox{sea quarks}\otimes(1+\cos\theta^2) \>.
\label{as0}
\end{equation}
Here $\theta$ denotes the angle between the electron (positron) and
the proton (antiproton) direction. The measurement of this angular
distribution\footnote{ At the CERN $Sp\overline{p}S$ collider energies
the contribution from  sea quark annihilation is negligible.} for the
$W$ boson by UA1 \cite{ua1} and UA2 \cite{ua2} established the $V-A$
coupling  and the spin-1 nature of the $W$ boson. For $Z$ boson
production at small $p_T^{}$, the  angular distribution can be
parametrized as
\begin{equation}
dN/d(\cos\theta) \sim 1+A_4\cos\theta+\cos^2\theta \>.
\end{equation}
The coefficient $A_4$ depends on the vector and axial-vector
couplings of the fermions to the $Z$ boson and is thus  sensitive to
$\sin^2\theta_W$.  Unfortunately, measurements of the $\cos\theta$
distributions  for gauge bosons produced at low $p_T^{}$ are only
sensitive to the trace of the density matrix elements of the  vector
bosons.

For gauge bosons produced with transverse momentum  [balanced by
additional gluons or quarks], the event plane spanned by the beam and
gauge boson momentum directions provides a convenient reference plane
for studying the angular distributions of the decay leptons. The
angular distributions are now sensitive  to non-diagonal elements of
the gauge boson density matrix elements. In leading order QCD [${\cal
O}(\alpha_s)$] the  angular distribution has the general form
\cite{cs}
\begin{eqnarray}
\sigma &\sim& (1+\cos\theta^2)
  + \frac{1}{2} \, A_{0} \, (1-3\cos^{2}\theta)
  +  A_{1}  \, \sin 2\theta \cos\phi \nonumber\\
 & &+ \frac{1}{2} \, A_{2}  \, \sin^{2}\theta\cos 2\phi
  +  A_{3}  \, \sin \theta \cos\phi
  +  A_{4}  \, \cos\theta  \>, \label{ai}
\end{eqnarray}
where $\theta$ and $\phi$ denote the polar and azimuthal angle of the
decay leptons in the gauge boson rest frame. Leading order (LO)
analytical  results for the coefficients $A_i$ for the decay of a  $W$
or $Z$ boson in the Collins--Soper \cite{cs} frame are discussed in
Refs.~\cite{chaichian,chiapetta,npb}. The complete
next-to-leading-order (NLO) corrections to the parity conserving
coefficients were calculated in Ref.~\cite{npb} and  were found to be
fairly small [less than 10\%]. This is because the  coefficients $A_i$
are  ratios of helicity cross sections  [see Eq.~(\ref{aintr})] and
the QCD corrections tend to   cancel in these ratios.  LO results for
the coefficients $A_0,A_1$, and $A_2$  for a virtual
photon\footnote{The parity-violating coefficients $A_3$ and $A_4$
vanish in this case.} $\gamma^\ast$  decaying into two leptons are
discussed in Ref.~\cite{tung}. An investigation  of the decay lepton
spectrum in the laboratory frame  can be found in Ref.~\cite{halzen}.

In this paper we present a numerical study of the  decay lepton
angular distribution for $W$ and $Z$ boson production at the Fermilab
Tevatron collider [$\sqrt{S} = 1.8$~TeV]. The calculation includes
acceptance cuts on the decay leptons. These cuts  are necessary  to
reduce the background from  low $p_T^{}$ jets which can fake
electrons. They are also necessary due to  the finite acceptance of
the detector. Uncertainties due to the  finite energy resolution of
the detector are simulated in the calculation by Gaussian smearing of
the  lepton four-momentum vectors. The energy resolution smearing has
a non-negligible effect only on the $\cos\theta$ distribution of the
lepton from $W$ decay. We show the transverse momentum [$p_T^{}(V)$]
dependence  of the coefficients $A_i$  and study the effect of the
acceptance cuts  and energy resolution smearing  on the $\phi$ and
$\cos\theta$ distributions in the Collins--Soper (CS) frame. The CS
frame has the unique advantage that the polar and azimuthal angles
$\theta$ and $\phi$ can be reconstructed   modulo a sign ambiguity in
$\cos\theta$ without information on the longitudinal momentum of the
neutrino.

Since the effects of the NLO corrections are small, it is sufficient
to use LO matrix elements in our Monte Carlo study. At NLO there are
three more angular coefficients  in Eq.~(\ref{ai}) which receive
contributions from the absorptive part of the  one-loop amplitudes. A
numerical analysis shows that  their  contribution  is  small [at the
level of 1\% at high  $p_T^{}$]  \cite{npb}, thus they will be
neglected in this analysis.

The measurements of the lepton distributions arising from the decay of
a  virtual photon $\gamma^*$ in nucleon-nucleon scattering presented
in Ref.~\cite{exp} appear to  disagree with the QCD improved parton
model. Several different nonperturbative effects have recently been
discussed in Ref.~\cite{nachtmann} to explain the discrepancy. It
would be interesting to test whether the angular distributions from
$W$ and $Z$ boson decay at the Tevatron are  in agreement  with
theoretical predictions.

The remainder of this paper is organized as follows:  the formalism
for describing the angular distributions is discussed in Sec.~II,
numerical results are presented in Sec.~III, and summary remarks are
given in Sec.~IV.

\section{ANGULAR DISTRIBUTIONS}

We consider the angular distribution of the leptons coming from the
leptonic decay of gauge bosons produced with non-zero transverse
momentum in high energy proton-antiproton collisions. For definiteness
we take
\begin{equation}
 p(P_{1}) + \bar{p}(P_{2}) \rightarrow W^{\pm}(Q) + X \rightarrow
 l^{\pm}(l_{1}) + \nu_l(l_{2}) + X \>,
\label{ppbarw}
\end{equation}
and
\begin{equation}
 p(P_{1}) + \bar{p}(P_{2}) \rightarrow Z(Q) + X \rightarrow
 l^{+}(l_{1}) + l^{-}(l_{2}) + X \>,
\label{ppbarz}
\end{equation}
where the quantities in parentheses denote the four-momenta of the
particles. In leading order QCD [${\cal O}(\alpha_s$)] the parton
subprocesses
\begin{eqnarray}
q + \bar{q} &\rightarrow& V + g \>, \nonumber \\
q + g       &\rightarrow& V + q  \>, \label{loprocess} \\
g + \bar{q} &\rightarrow& V + \bar{q} \>, \nonumber
\end{eqnarray}
contribute to high $p_{T}^{}$ gauge boson production.

In the parton model the hadronic cross section is obtained by folding
the hard parton level cross section with the respective parton
densities:
\begin{equation}
\frac{d\sigma^{h_{1}h_{2}}}{ dp_{T}^{2}\,dy\,d\Omega^{\ast} } =
\sum_{a,b}
\int\,\,dx_{1}\,dx_{2}\,
f_{a}^{h_{1}}(x_{1},\mu_F^{2})\,
f_{b}^{h_{2}}(x_{2},\mu_F^{2})
\,
\frac{{s}\,d{\hat{\sigma}}_{ab}}{ d{t}\,d{u}\,d\Omega^{\ast}}\,
\left( x_{1}P_{1},x_{2}P_{2},\as(\mu_R^{2}) \right) \>,
\label{wqhad}
\end{equation}
where the sum is over $a,b=q, \bar{q}, g$. $\,f_{a}^{h}(x,M^{2})$ is
the probability density for finding parton $a$ with momentum fraction
$x$ in hadron $h$ when it is probed at scale $\mu_F^{2}$. The parton
level  cross section for the processes in Eq.~(\ref{loprocess}) are
denoted by $d{\hat{\sigma}}_{ab}$. Denoting hadron level and parton
level quantities by upper and lower case characters, respectively,
the hadron and parton level Mandelstam variables are  defined by
\begin{equation}
S = (P_1 + P_2)^2 \>, \qquad T = (P_1 - Q)^2 \>, \qquad U = (P_2 - Q)^2 \>,
\end{equation}
and
\begin{equation}
  \begin{array}{lclcl}
\label{kleinmandeldef}
s&=&(p_{1}+p_{2})^{2}&=&x_{1}x_{2}S \>, \\[2mm]
t&=&(p_{1}-Q)^{2}    &=&x_{1}(T-\qq)+\qq \>, \\[2mm]
u&=&(p_{2}-Q)^{2}    &=&x_{2}(U-\qq)+\qq \>,
  \end{array}
\end{equation}
where $p_1 = x_1 P_1$ and $p_2 = x_1 P_2$. The rapidity $y$ of the
gauge boson in the laboratory frame can be written
\begin{equation}
y=\frac{1}{2}\log\left(\frac{\qq-U}{\qq-T}\right) \>,
\end{equation}
and the transverse momentum $p_T^{}$ of the gauge boson is related to
the Mandelstam variables via
\begin{equation}
p_{T}^{2}=\frac{(\qq-U)(\qq-T)}{S}-\qq = \frac{ut}{s} \>.
\end{equation}
The angles $\theta$ and  $\phi$ in $d\Omega^{\ast} =
d\cos\theta\,d\phi$ are the polar and azimuthal decay angles of the
leptons in the gauge boson rest frame measured  with respect to a
coordinate system to be described later.

Technically, the lepton-hadron correlations are described by the
contraction of the lepton tensor $L_{\mu\nu}$ with the hadron tensor
$H^{\mu\nu}$, where $L_{\mu\nu}$ acts as an analyzer of the  gauge
boson polarization. The angular dependence in Eq.~(\ref{wqhad}) can be
extracted by introducing  helicity cross sections corresponding to the
non-zero combinations of the polarization density matrix elements
\begin{equation}
H_{m m^{\prime}} = \epsilon_{\mu}^{\ast}(m) \, H^{\mu\nu} \,
\epsilon_{\nu}(m^{\prime}) \>,
\end{equation}
where $m,m^{\prime}=+1,0,-1$ and
\begin{equation}
    \begin{array}{ll}
\epsilon_{\mu}(\pm 1)&=\frac{1}{\sqrt{2}}\,(0;\pm1,-i,0) \>, \\[3mm]
\epsilon_{\mu}(0)    &=(0;0,0,1) \>,
    \end{array}
\hspace{5mm}
\label{polvekdef}
\end{equation}
are the polarization vectors for the gauge boson defined with respect
to the chosen gauge boson rest frame. The angular dependance of the
differential cross section can be written [see Ref.~\cite{npb} for
details]
\begin{eqnarray}
\frac{16\pi}{3}\frac{d{\sigma}}{dp_T^2\,dy\,d\cos\theta\,d\phi} =
&\phantom{+}& \frac{d{\sigma}^{U+L}}{dp_T^2\,dy}\,(1+\cos^2\theta)
\,+\, \frac{d{\sigma}^{L}}{dp_T^2\,dy}\,(1-3\cos^2\theta)\nonumber\\[2mm]
&+& \frac{{d\sigma}^{T}}{dp_T^2\,dy}\,2\sin^2\theta\cos2\phi
\,+\, \frac{d{\sigma}^{I}}{dp_T^2\,dy}\,
2\sqrt{2}\sin2\theta\cos\phi \label{hadronwinkel} \\[2mm]
&+&
 \frac{d{\sigma}^{A}}{dp_T^2\,dy}\,
4\sqrt{2}\sin\theta\cos\phi
\,+\,
\frac{{d\sigma}^{P}}{dp_T^2\,dy}\,2\cos\theta \>, \nonumber
\end{eqnarray}
where
\begin{equation}
    \begin{array}{lllll}
\sigma^{U+L} & \sim \hspace{3mm}H_{00}+ H_{++} + H_{--} \>, &\hspace{5mm}&
        &% = H_{11}+H_{22}+H_{33}
          \\[3mm]
\sigma^{L} & \sim \hspace{3mm}H_{00} \>,         & \hspace{5mm}&
        &% = H_{33}
        \\[3mm]
\sigma^{T} & \sim\hspace{3mm} \frac{1}{2}( H_{+-}+H_{-+}) \>,  &\hspace{5mm}&
        & %= \frac{1}{2}(H_{22}-H_{11})
        \\[3mm]
\sigma^{I} & \sim \hspace{3mm}\frac{1}{4}
           \left(H_{+0}+H_{0+}-H_{-0}-H_{0-} \right) \>, &\hspace{5mm}&
      &% = -\frac{1}{2\sqrt{2}}(H_{31}+H_{13})
        \\[3mm]
\sigma^{P} & \sim \hspace{3mm}H_{++} - H_{--} \>, & \hspace{5mm}&
            &%=
           % -i(H_{12}-H_{21})
            \\[3mm]
\sigma^{A} & \sim \hspace{3mm}
          \frac{1}{4}\left( H_{+0} + H_{0+} +H_{-0}+H_{0-} \right) \>.
          &\hspace{5mm}&&       %=
          %-\frac{i}{2\sqrt{2}}(H_{23}-H_{32})
        \\[3mm]
    \end{array}
\label{hidef1}
  \end{equation}
The unpolarized differential production cross section is denoted by
${\sigma}^{U+L}$ whereas ${\sigma}^{L,T,I,P,A}$ characterize the
polarization of the gauge boson, {\it e.g.}, the cross section for
longitudinally polarized gauge bosons is denoted by ${\sigma}^L$, the
transverse-longitudinal interference cross section by ${\sigma}^I$,
the transverse interference cross section by ${\sigma}^T$, {\it etc}
(all with respect to the $z$-axis of the chosen lepton pair rest
frame; for details see appendix C of Ref.~\cite{npb}). At NLO there
are three more ``T-odd'' angular coefficients in
Eq.~(\ref{hadronwinkel}) for $W$ and $Z$ production. However, their
numerical contribution is  small \cite{npb,hagiwara} and we will neglect them
in the present analysis. The helicity cross sections $\sigma^{\alpha}$
contain the following coupling coefficients:
\begin{eqnarray}
\sigma^{U+L,L,T,I}&\sim&(v_{l}^{2}+a_{l}^{2})(v_{q}^{2}+a_{q}^{2}) \>, \\
\sigma^{P,A} &\sim& v_{l}\, a_{l}\, v_{q}\, a_{q} \label{va}\>,
\end{eqnarray}
where $v_{q}(v_{l})$ and $a_{q}(a_{l})$ denote the vector and axial
vector coupling of the gauge boson to the quark (lepton).

The hadronic  helicity cross sections
$\frac{d{\sigma}^{\alpha}}{dp_T^2\,dy}$ in Eq.~(\ref{hadronwinkel})
are obtained by convoluting the partonic helicity  cross sections with
the parton  densities:
\begin{equation}
\frac{d\sigma^{\alpha}}{ dp_{T}^{2}\,dy } =
\int\,\,dx_{1}\, dx_{2}\,
f^{h_{1}}(x_{1},\mu_F^{2})\,
f^{h_{2}}(x_{2},\mu_F^{2})\,
\frac{{s}\,d{\hat{\sigma}^\alpha}}{ d{t}\,d{u}} \>.
\label{wqalphahad}
\end{equation}
The helicity cross sections $\hat{\sigma}^{\alpha}$ are dependent on
the choice of the $z$-axis in the rest frame of the gauge boson and
are explicitly given  in Ref.~\cite{npb} for the Collins--Soper
\cite{cs} frame. In this frame the $z$-axis bisects the angle between
${\vec{P}_{1}}$ and ${-\vec{P}_{2}}$:
\begin{equation}
          \begin{array}{ll}
&{\vec{P}_{1}} = E_{1}\,(\sin\gamma_{CS}^{},0, \cos\gamma_{CS}^{}) \>, \\[1mm]
&{\vec{P}_{2}} = E_{2}\,(\sin\gamma_{CS}^{},0,-\cos\gamma_{CS}^{}) \>,
          \end{array}
\end{equation}
with
\begin{eqnarray}
\cos\gamma_{CS}^{} &=& \left( \frac{\qq S}{(T-\qq)(U-\qq)} \right)^{1/2}
                    =  \left( \frac{\qq}{\qq+p_T^2} \right)^{1/2} \>,
\label{winkelcsdef} \\[1mm]
\sin\gamma_{CS}^{} &=& -\sqrt{1-\cos^2\gamma_{CS}^{}} \>,\\[1mm]
E_1&=&\frac{Q^2-T}{2\sqrt{Q^2}}\>,
\hspace{1cm}
E_2=\frac{Q^2-U}{2\sqrt{Q^2}}\>.
\end{eqnarray}
In the laboratory frame, the $z$-direction is defined by the proton
momentum and the $x$-direction is defined by the transverse momentum
of the gauge boson. The Lorentz transformation matrix for the  boost
from the laboratory frame to the CS frame is given by
\begin{equation}
\left(
\begin{array}{c}
E\\[3mm]
p_x\\[3mm]
p_y\\[3mm]
p_z\\[3mm]
\end{array}
\right)_{CS}
= \
\left(
\begin{array}{cccc}
\frac{Q_0}{\sqrt{Q^2}}\hspace{3mm} & -\frac{p_T^{}}{\sqrt{Q^2}}   &
\hspace{3mm}0\hspace{3mm}       & -\frac{Q_3}{\sqrt{Q^2}}        \\[3mm]
-\frac{p_T^{}\, Q_0}{\sqrt{Q^2}\,X_T} & \frac{X_T}{\sqrt{Q^2}}    &
0                      & \frac{p_T^{}\,Q_3}{\sqrt{Q^2}\,X_T}\\[3mm]
0                      & 0                                     &
1                      & 0                               \\[3mm]
-\frac{Q_3}{X_T}       & 0                                     &
0                      & \frac{Q_0}{X_T}\\[3mm]
\end{array}
\right)
\
\left(
\begin{array}{c}
E\\[3mm]
p_x\\[3mm]
p_y\\[3mm]
p_z\\[3mm]
\end{array}
\right)_{lab}
\label{EQ:BOOST}
\end{equation}
where
\begin{equation}
X_T=\sqrt{Q^2+p_T^2} \>.
\end{equation}

Concerning the leptonic decay modes, $W$ bosons have the disadvantage
that the decay kinematics cannot be completely reconstructed due to
the unobservability of the neutrino. The transverse components of the
neutrino's momentum can be approximated from transverse momentum
balancing in calorimetric experiments, however, the longitudinal
component can only be obtained up to a twofold ambiguity. By choosing
the Collins--Soper frame \cite{cs}, the polar angle ($\theta$) and the
azimuthal angle ($\phi$) of the  charged lepton can be reconstructed
modulo a sign ambiguity in $\cos\theta$, thereby enabling the
measurement of  four [$\sigma^{U+L,L,T,A}$] out of the six helicity
cross sections in Eq.~(\ref{hadronwinkel}) \cite{npb}. For $Z$ boson
production and decay, $\phi$ and $\cos\theta$ can  be determined
without ambiguity.

Let us briefly review how $\theta$ and $\phi$  can be obtained from
measured quantities in $W$ boson production and decay. Denote the
charged lepton's four-momentum in the laboratory frame by
$p_l=(E_l,p_{lx},p_{ly},p_{lz})$ where $p_{lz}$ and $p_{lx}$ are the
lepton's momentum components parallel and perpendicular to the proton
direction in  the event plane, {\it i.e.}, the plane spanned by the
proton and the $W$ boson with $p_T^{} > 0$. The lepton's four-momentum
in the CS frame, $p_l^{CS}$, can be obtained by Lorentz-boosting $p_l$
from the laboratory frame to the CS frame. One obtains
\begin{eqnarray}
 p_{lx}^{CS} & = & \frac{1}{2} \frac{M_W}{\sqrt{M_W^2 + p_T^2}}
  \; (2\, p_{lx} - p_T^{}) \>,
\nonumber       \\
 p_{ly}^{CS} & = &  p_{ly} \>,
\label{plcs}     \\
 p_{lz}^{CS} & = & \pm \frac{M_W}{2}
    \left(1 - \frac{\left(p_{lx}^{CS}\right)^2 + \left(p_{ly}^{CS}\right)^2}
                   {M_W^2/4} \right)^{1/2} \>,          \nonumber
\end{eqnarray}
where the $W$-mass constraint has been imposed on the lepton-neutrino
system. The $\pm$~signs correspond to the two solutions for the
longitudinal momentum of the neutrino. The transverse component
$p_{lx}^{CS}$ is uniquely determined by the measurable laboratory
frame quantities\footnote{ Note that there is a misprint in the
definition of $p_T^{}$ in Ref.~\cite{npb}.} $p_{lx}$ and $p_T^{}=
p_{lx} + p_{\nu x}$. The CS frame is thus  unique in the sense  that
in this frame the lepton's transverse momentum is independent of the
unmeasured longitudinal momentum of the neutrino. On the other hand
note that the longitudinal component $p_{lz}^{CS}$ is determined only
up to a sign. The angles $\theta$ and $\phi$ are obtained from the
charged  lepton's momentum components in Eq.~(\ref{plcs})  via $\phi =
\arctan{\left(p_{ly}^{CS}/p_{lx}^{CS}\right)}$ and $\theta =
\arctan{\sqrt{(p_{lx}^{CS})^2 + (p_{ly}^{CS})^2} /p_{lz}^{CS}}$. The
twofold ambiguity in the reconstruction of the lepton's longitudinal
momenta in the CS frame translates into an ambiguity $\theta
\leftrightarrow \pi - \theta$ in the polar angle, while $\phi$ is
completely determined. This implies that only the helicity cross
sections $\sigma^{U+L},\sigma^{L},\sigma^{T},$ and $\sigma^{A}$  in
Eq.~(\ref{hadronwinkel}) can be determined in a $W$ boson production
experiment.   For $Z$ boson production, all six of the angular
coefficients in Eq.~(\ref{hadronwinkel}) can be determined.

Introducing the standard angular coefficients \cite{cs}
%
%
%\begin{equation}
$$
A_{0}=\frac{2\,\, d\sigma^{L}}{d\sigma^{U+L}} \>, \hspace{1cm}
A_{1}=\frac{2\sqrt{2}\,\, d\sigma^{I}}{d\sigma^{U+L}} \>, \hspace{1cm}
A_{2}=\frac{4 \,\,d\sigma^{T}}{d\sigma^{U+L}} \>, \hspace{1cm}
$$
%\end{equation}
\begin{equation}
A_{3}=\frac{4\sqrt{2}\,\, d\sigma^{A}}{d\sigma^{U+L}} \>, \hspace{1cm}
A_{4}=\frac{2 \,\,d\sigma^{P}}{d\sigma^{U+L}} \>, \hspace{1cm}
\label{aintr}
\end{equation}
the angular distribution in Eq.~(\ref{hadronwinkel}) can be
conveniently written
\begin{eqnarray}
\frac {d\sigma}{d p_{T}^{2}\,dy\, d\cos\theta \,d\phi}
&=&  \frac{3}{16\pi}\,
\frac{d\sigma^{U+L}}{ d p_{T}^{2}\,dy}\,\,
           \,\left[  (1+\cos^{2}\theta) \nonumber
               +\,\, \frac{1}{2}A_{0} \,\,\, (1-3\cos^{2}\theta)
 \right.\\[2mm]
&&   \left.
\hspace{1.5cm} +  \,\,   A_{1}  \,\,\,\sin 2\theta \cos\phi \,\,
\,\, + \,\,   \frac{1}{2}A_{2}  \,\,\,\sin^{2}\theta\cos 2\phi\nonumber
  \right. \nonumber\\
&&   \left.
\hspace{1.5cm} +  \,\,   A_{3}  \,\,\,\sin \theta \cos\phi \,\,
\,\, + \,\,   A_{4}  \,\,\,\cos\theta
\,\,  \right] \>.
  \label{ang}
  \end{eqnarray}
Integrating the angular distribution in Eq.~(\ref{ang}) over the
azimuthal angle $\phi$ yields
\begin{equation}
\frac{d\sigma}{dp_{T}^{2}\,dy\,d\cos\theta}=C\,
                  (1+\alpha_{1}\cos\theta+\alpha_{2}
\cos^{2}\theta) \>,
\label{alphadef}
\end{equation}
where
\begin{equation}
C= \,\frac{3}{8}\,
\frac{d\sigma^{U+L}}{dp_{T}^{2}\,dy}
\left[1+\frac{A_{0}}{2}\right]\>, \hspace{1cm}
\alpha_{1}=\frac{2\,A_{4}}{2+A_{0}} \>, \hspace{1cm}
\alpha_{2}=\frac{2-3A_{0}}{2+A_{0}} \>.
\label{alphaidef}
\end{equation}
Integrating Eq.~(\ref{ang}) over the polar angle $\theta$ yields
\begin{equation}
\frac{d\sigma}{dp_{T}^{2}\,dy\,d\phi}=
\,\frac{1}{2\pi}\,
\frac{d\sigma^{U+L}}{dp_{T}^{2}\,dy} \,
         (1+\beta_{1}\cos\phi+\beta_{2}\cos 2\phi ) \>,
%           +\beta_{3}\sin\phi+\beta_{4}\sin 2\phi)
\label{betadef}
\end{equation}
where
\begin{equation}
\beta_{1}=\frac{3\pi}{16}A_{3} \>, \hspace{1cm}
\beta_{2}=\frac{A_{2}}{4} \>. \hspace{1cm}
%\beta_{3}=\frac{3}{4}A_{7}\hspace{1cm}
%\beta_{4}=\frac{A_{5}}{2}\hspace{1cm}
\label{betaidef}
\end{equation}

Before discussing numerical results, let us briefly discuss  a
possible strategy for extracting the angular coefficients. By taking
moments with respect to an appropriate product of trigonometric
functions it is possible to disentangle the coefficients $A_i$. A
convenient definition of the moments is
\begin{equation}
\langle m \rangle =
\frac{\int d\sigma(p_T^{},y,\theta,\phi)\,\, m \,\,
{d\cos\theta}\,{d\phi}}{
\int d\sigma(p_T^{},y,\theta,\phi)\,
\,{d\cos\theta}\,{d\phi}} \>,
\label{moment}
\end{equation}
which leads to the following results:
\begin{eqnarray}
\langle 1 \rangle &=& 1 \>, \\[2mm]
\langle \frac{1}{2}(1-3\cos^2\theta ) \rangle &=&
\frac{3}{20}\,\,\left( A_0-\frac{2}{3}\right) \>, \\[2mm]
\langle \sin 2\theta\cos\phi  \rangle &=& \frac{1}{5} \,\,A_1 \>, \\[2mm]
\langle \sin^2\theta\cos2\phi  \rangle &=&\frac{1}{10}\,\, A_2 \>, \\[2mm]
\langle \sin\theta \cos\phi \rangle &=& \frac{1}{4}\,\, A_3 \>, \\[2mm]
\langle \cos\theta  \rangle &=& \frac{1}{4}\,\,A_4 \>.
\end{eqnarray}

\section{NUMERICAL RESULTS}

In this section numerical results are presented for high $p_T^{}$
production and leptonic decay of $W$ and $Z$ bosons at the Tevatron
collider center of mass energy [$\sqrt{S} = 1.8$~TeV]. The numerical
results have been obtained using the MRS set $D_{-}^{\prime}$
\cite{MRS} parton distribution functions  with $\Lms{4}= 215$~MeV and
the one-loop formula for $\alpha_s$. The renormalization scale
$\mu_R^2$ and the factorization scale $\mu_F^2$ in Eq.~(\ref{wqhad})
have been taken to be $\mu_R^2 = \mu_F^2 = M_V^2+p_T^2(V)$, where
$M_V^{}$ and $p_T^{}(V)$ are the mass and transverse momentum,
respectively, of the gauge boson.

We begin with numerical results for the coefficients $A_i$ in
Eq.~(\ref{ang}).  These coefficients have been extracted from  the
Monte Carlo program by using the moments defined in
Eq.~(\ref{moment}). Figure~1(a) shows the coefficients  $A_0, A_2,$
and $A_3$ in the CS frame as functions of  $p_T^{}(W)$ for $W$
production. The angles $\theta$ and $\phi$ are the decay angles of the
charged lepton in the CS frame.  The results are identical for $W^+$
and $W^-$. The lepton's four-momentum in the CS frame was obtained
using  Eq.~(\ref{plcs}). Note that because of the ambiguity
$\theta\leftrightarrow\pi-\theta$ in the polar angle reconstruction in
the CS frame, the coefficients $A_1$ and $A_4$    are not observable
in a $W$ boson production experiment. However, all of the $A_i$
coefficients are observable for $Z$ boson production, and the
numerical results are shown in Fig.~1(b), where $\theta$ and $\phi$
are the angles of the negatively charged lepton. No acceptance cuts
have been applied to the leptons.

The coefficients $A_0$ and $A_2$ are increasing functions  of
$p_T^{}(V)$ and the deviations from the lowest order expectation of
Eq.~(\ref{as0}) [$A_0=A_2=0$] are quite large, even at modest values
of $p_T^{}(V)$, {\it i.e.}, for $p_T^{}(V) \approx 20-50$~GeV.  It has
been noted in  Ref.~\cite{tung} that these coefficients are  exactly
equal in LO. This is no longer true in NLO, but the corrections to
$A_0$ and $A_2$ are fairly small \cite{npb}. It has been shown in
Ref.~\cite{arn} that the $O(\alpha_s)$ relation $A_0=A_2$ is peculiar
to a vector gluon theory. For a scalar gluon this relation is badly
broken, {\it e.g.}, $A_0-A_2\approx -2$ in the Gottfried-Jackson frame
at the CERN $Sp\bar{p}S$ energies. In was in fact the study of this
relation in $W$ decays by UA1  which fixed the spin of the gluon
\cite{ua1}. NLO numerical results for $A_0$ and $A_2$ for $W$ decay in
the Gottfried-Jackson frame are given  in Ref.~\cite{m91}  and the
deviations from $A_0=A_2$ are  found to be small. NLO numerical
results for $A_0$ and $A_2$ in the CS frame are presented in
Refs.~\cite{npb,mks91}.

The deviation of $A_1$ and $A_3$ from zero is much smaller, even at
large $p_T^{}(V)$. This is a special effect of the CS frame. In this
frame   the $q\bar{q}$ contribution in Eq.~(\ref{loprocess}) to
$A_{1}$ and  $A_{3}$ is antisymmetric under the interchange of $x_{1}$
and $x_{2}$; this can be explicitly seen in the corresponding matrix
elements in Eq.~(24) of Ref.~\cite{npb}. The $qg$ process gives a
sizable contribution  to $A_3$ only for $W$ boson production, and a
measurement of $A_{3}$  could   in principle provide  constraints on
the gluon distribution function\footnote{Note that   $A_3$ has the
wrong sign  in the  figures in Ref.~\cite{npb}}. However, as we will
see later, experimental cuts introduce additional complicated angular
effects and the resulting data sample can no longer be described by
the simple angular distribution in Eq.~(\ref{ang}). Futhermore, the
extraction of the different polarization cross sections via the
moments in Eq.~(\ref{moment}) becomes problematic. Note that $A_4$ in
Fig.~1(b) is proportional to $v^Z_{l}\, a^Z_{l}\, v^Z_{q}\, a^Z_{q}$
and is therefore sensitive to the Weinberg angle $\theta_W$. This
determination would  be totally independent of the $W$ and $Z$ boson
masses. However,    a precise determination of $\sin^2\theta_W$ by
this method is also limited by the problem of extracting $A_4$ with
sufficient  accuracy when experimental cuts are applied to the decay
leptons [see the discussion below].

To give a feeling for the $\phi$ and $\cos\theta$ distributions of the
decay leptons in the CS frame we show numerical results for the
coefficients $\alpha_1,\alpha_2$  and $\beta_1, \beta_2$ [see
Eqs.~(\ref{alphadef}) and (\ref{betadef})] as a function of $p_T^{}(V)$
for the $W$ and $Z$ boson in Fig.~2.  As in Fig.~1, no cuts have been
applied to the leptons. The coefficients are again very sensitive to
the transverse momentum of the gauge boson. In the remainder of this
paper we investigate the feasibility of using this dependence to test
the polarization  properties of the $W$ and $Z$ bosons at the
Tevatron.

We want  to point out that we are not in agreement with the
theoretical predictions  for the angular coefficients of the $Z$ boson
presented in Ref.~\cite{chiapetta} where the ``soft gluon
resummation'' formalism was used for the  integrated cross
section\footnote{The integrated cross  section is essentially the
total cross section after integration over  $\theta$ and $\phi$, {\it
i.e.}, $\sigma^{U+L}$ in Eq.~(\ref{hadronwinkel}).} but not for the
individual helicity cross sections which are responsible for the
angular distributions.  Even for  gauge boson transverse momenta
larger than $p_T^{} > 20$~GeV, where the angular coefficients should
be reliably predicted by fixed order perturbative QCD, our predictions
differ dramatically from the results in Ref.~\cite{chiapetta}, for
example, by a factor five for the coefficient $A_0$ in Eq.~(\ref{ang})
at $p_T^{}(Z)=60$~GeV. The complete NLO calculation in Ref.~\cite{npb}
for the helicity cross sections shows that the deviations from the LO
result are much smaller than the predictions in Ref.~\cite{chiapetta}.

Figures~3--13 show the normalized  $\phi$ and $\cos\theta$
distributions for the leptons from the decay of $W$ and $Z$ bosons for
four bins in the transverse momentum of the gauge boson. To
demonstrate the effects of acceptance cuts, results are shown first
without cuts and then with typical acceptance cuts  imposed on the
leptons. The acceptance cuts are necessary  to reduce the background
from  low $p_T^{}$ jets which can fake electrons. They are also needed
due to the finite acceptance of the detector. Measurements of the
charged lepton momentum  and the missing transverse momentum
$p\llap/_T^{}$  have inherent uncertainties due to the finite energy
resolution of the detector.  These uncertainties have been simulated
in our calculation by Gaussian smearing of the charged lepton and
neutrino four-momentum vectors with standard deviation $\sigma$.  The
numerical results presented here were made using $\sigma$ values based
on the CDF specifications \cite{SMEARING}. The energy resolution
smearing has a non-negligible effect only on the $\cos\theta$
distribution of the charged lepton from $W$ decay.

Figure~3 shows the $\phi$ and $\cos\theta$ distributions for the $W$
boson  without cuts or smearing. The curves in  Fig.~3 can be obtained
from  $\alpha_2$, $\beta_1$, and $\beta_2$ in Fig.~2. For example,  in
the lowest $p_T^{}(W)$ bin [solid curve], the $\phi$ distribution is
almost flat [$\beta_1,\beta_2 \approx 0$] and the $\cos\theta$
distribution is approximately $1+\cos\theta^2$ [$\alpha_2 \approx 1$].
In the highest $p_T(W)$ bin [dot-dashed curve], one observes the
$\phi$ dependence [$\beta_1,\beta_2\neq 0$] resulting from the
non-diagonal elements of the spin-density matrix of the $W$ boson,
whereas the corresponding $\cos\theta$ distribution is almost flat
[$\alpha_2 \approx 0$]. Note that if the $W$ boson was to decay
isotropically, the $\phi$ and $\cos\theta$ distributions would both be
flat.

Figure~4 shows the $\phi$ and $\cos\theta$ distributions for the $W$
boson for the same bins in $p_T^{}(W)$ as in  Fig.~3, but with energy
resolution smearing and the cuts
\begin{equation}
p_T^{}(l) > 25\ \mbox{GeV},
\hspace{10mm}
p\llap/_T^{} > 25\ \mbox{GeV},
\hspace{10mm}
y(l) < 1.
\label{wcuts}
\end{equation}
Only 33\% of the events pass the cuts in Eq.~(\ref{wcuts}). The cuts
have a dramatic effect on the shapes of the distributions. The shapes
of the distributions are now governed by the kinematics of the
surviving events.  The cuts, which are applied in the laboratory
frame, introduce a strong $\phi$ dependence. The ``kinematical''
$\phi$ distribution in Fig.~4(a) is  very different from the
``dynamical'' $\phi$ distribution in Fig.~3(a).   The only remaining
vestiges of the polarization  effects in the $\phi$ distribution are
the dips in the high $p_T^{}(W)$ curve [dot-dashed curve] at $\phi =
90^\circ,\, 180^\circ,$ and $270^\circ$.

The $\cos\theta$ distributions with cuts in Fig.~4(b) are very
different from the corresponding  results without cuts or smearing in
Fig.~3(b). In order to separate the effects of smearing from the
effects of cuts, the $\cos\theta$ distribution is shown in Fig.~5 with
the cuts of Eq.~(\ref{wcuts}), but without smearing. The main effect
of the cuts is to remove events with  $\cos\theta > 0.5$ for small
values of $p_T^{}(W)$, causing  the curves to drop to zero as
$\cos\theta\rightarrow 1$  [see the solid and dashed curves in
Fig.~5]. Comparing Figs.~4(b) and 5, we see that the smearing of the
lepton momentum and the missing transverse momentum has a large effect
on the $\cos\theta$ distribution in the small $\cos\theta$ region.
This behavior of the smeared curves can be traced to the expression
for $p_{lz}^{CS}$ in Eq.~(\ref{plcs}).   Small values of $\cos\theta$
correspond to small values of  $p_{lz}^{CS}$. When events with small
values of $\cos\theta$ are smeared, the argument of the square root
function in the expression  for $p_{lz}^{CS}$ often becomes negative,
in which case the argument is set equal to zero in the Monte Carlo
program.  Thus many events with small values of $\cos\theta$ are
smeared such that they end up in the $\cos \theta = 0$ bin, which is
not shown here since it is off the vertical scale. When cuts and
smearing are combined, as in Fig.~4(b), very little  polarization
dependence is left in the $\cos\theta$ distribution for the $W$ boson.

We have also analyzed  the effect of the cuts separately by using  the
correct matrix element for $W$ boson production, but with isotropic
decay of the $W$ boson, {\it i.e.}, neglecting spin correlations
between $W$ production and decay. The angular distributions in this
case are very similar to the ones shown in Fig.~4 for the full matrix
element;  the remnant polarization effects discussed in Fig.~4 are of
course absent. In Fig.~6 we show ratios of the $\phi$ and $\cos\theta$
distributions for the same bins in $p_T^{}(W)$ as in Figs.~3--5; the
distribution with full polarization has been divided by the
distribution obtained with isotropic decay of the $W$ boson. Cuts and
smearing are included in both cases. The large effects from the cuts
and smearing are expected to almost cancel in this ratio. In fact, we
nearly recover the $\phi$ and $\cos\theta$ dependence of Fig.~3 which
contains no cuts or smearing. Therefore, to regain sensitivity to the
polarization effects in the presence of large kinematic cuts, we
propose to divide the experimental distributions by the Monte Carlo
distributions obtained using isotropic gauge boson decay.

Figure~7 again shows the $\phi$ and $\cos\theta$ distributions  for
the $W$ boson with energy resolution smearing, but now  with the
looser cuts
\begin{equation}
p_T^{}(l) > 15\ \mbox{GeV},
\hspace{10mm}
p\llap/_T^{} > 15\ \mbox{GeV},
\hspace{10mm}
y(l) < 2.5 \>.
\label{wcutslow}
\end{equation}
The results now display more of the polarization effects seen in
Fig.~3. The dips in the $\phi$ distribution in the high $p_T(W)$ curve
[dot-dashed curve] at $\phi=90^\circ$ and $270^\circ$ in Fig.~7(a)
are much more distinctive than in Fig.~4(a). Dynamical effects in the
$\cos\theta$ distribution [see Fig.~3(b)] can still be observed in
Fig.~7(b) for $\cos\theta$ values in the range $0.3<\cos\theta<0.8$.
Figure~8 shows  the ratio of the ``polarized'' $\phi$ and $\cos\theta$
distributions to the corresponding distributions with isotropic
leptonic decay for the cuts in Eq.~(\ref{wcutslow}). Most of the
polarization dependence seen in Fig.~3 is retained in this ratio.

The $\phi$ and $\cos\theta$ distributions of the leptons from $Z$
boson decay are shown in Figs.~9--13 for four bins in $p_T^{}(Z)$. For
$Z$ bosons, the lepton momenta in the CS frame  can be obtained from
the measured lepton momenta in the laboratory frame by applying the
boost matrix given in Eq.~(\ref{EQ:BOOST}). The $\phi$ and
$\cos\theta$ distributions of the negatively charged lepton are shown
without cuts in Fig.~9.  Again, there are large differences between
the different $p_T^{}(Z)$ bins. If the electric charge of the lepton
can be determined, then the coefficient $A_4$ [or $\alpha_1$] can in
principle be measured and  the $\cos\theta$ distribution in Fig.~9(b)
is asymmetric about $\cos \theta = 0$. However, if the lepton's
electric charge can not be determined,  the term linear in
$\cos\theta$ will be averaged out, and the resulting $\cos\theta$
distribution will be symmetric.

Figure~10 shows the $\phi$ and $\cos\theta$ distributions for the $Z$
boson  using  the same bins in $p_T^{}(Z)$  as in Fig.~9, but now with
energy resolution smearing and the cuts
\begin{equation}
p_T^{}(l) > 25\ \mbox{GeV},
\hspace{10mm}
y(l)<1.
\label{zcuts}
\end{equation}
Note that these cuts are more stringent than the cuts for the $W$
boson in Eq.~(\ref{wcuts}) since now  both leptons must be in the
central rapidity region. The effect of the energy resolution smearing
is negligible for the $Z$ boson. The cuts introduce a $\phi$
dependence similar to that discussed for the $W$ boson case in
Fig.~4(a).  Some polarization effects are still observable in the
$\phi$ distribution in the high $p_T^{}(Z)$ bins, for example,  the
dip in the dot-dashed curve and the flat behavior in the dotted curve
around $\phi=90^\circ$ and $270^\circ$ in Fig.~10(a) are due to
polarization effects.

The cuts in Eq.~(\ref{zcuts}) have an even more dramatic effect  on
the $\cos\theta$ distribution. Since nearly all events with
$|\cos\theta|>0.6$ are rejected, almost all sensitivity to the
asymmetry in Fig.~9(b) is lost. Futhermore,  the curves in Fig.~10(b)
are fairly similar in shape. Therefore it may be very difficult to
observe any polarization effect from the $Z$ boson in these
distributions when the  cuts in Eq.~(\ref{zcuts}) are imposed.
However, by forming the ratio of  the $\phi$ and $\cos\theta$
distributions with full polarization to the corresponding
distributions obtained with isotropic $Z$ decay, most of the
polarization dependence is recovered; see Fig.~11.

Since the experimental signature of the $Z$ boson decay is much
cleaner than that for the $W$ boson, Fig.~12 shows the $\phi$ and
$\cos\theta$ distributions for the $Z$ boson for the looser cuts
\begin{equation}
p_T^{}(l) > 10\ \mbox{GeV},
\hspace{10mm}
y(l) < 2.5 \, .
\label{zzcuts}
\end{equation}
The large dips in the $\phi$ distribution in the high $p_T^{}(Z)$ bin
[dot-dashed curve] at $\phi=90^\circ$ and $270^\circ$  are due to
polarization effects [compare Fig.~12(a) to Fig.~9(a)]. The curves in
the central region of $\cos\theta$ in  Fig.~12(b) now clearly exhibit
the polarization effects  seen in the corresponding curves in
Fig.~9(b). Finally, Fig.~13 shows the ratio of the ``polarized''
$\phi$ and $\cos\theta$ distributions to the corresponding
distributions obtained with isotropic leptonic decay for the $Z$ boson
for the cuts in Eq.~(\ref{zzcuts}).  The ratios once again  contain
most of the polarization dependence seen in Fig.~9.

\section{SUMMARY}

The polar and azimuthal angular distributions of the lepton pair
arising from the decay of a $W$ or $Z$ boson produced at high
transverse momentum in hadronic collisions have been discussed.  In
the absence of cuts on the final state leptons, the general structure
of the lepton angular distribution in the gauge boson rest frame is
determined by the gauge boson polarization.  At ${\cal O}(\alpha_s)$
in perturbative QCD, the structure is described by six helicity cross
sections, which are functions of the transverse momentum and rapidity
of the gauge boson.

We have studied the angular distributions of the leptonic decay
products of high $p_T^{}$ gauge bosons when acceptance cuts and energy
resolution smearing are applied to the leptons.   When acceptance cuts
are imposed on the leptons, the shapes of the lepton angular
distributions are dominated by kinematic effects and the residual
dynamical effects from the gauge boson polarization are small.  The
kinematic effects become more dominate as the cuts become more
stringent.  Energy resolution smearing has a significant effect only
on the $\cos\theta$ distribution from $W$ decay.  The large smearing
effect in this distribution is a consequence of the undetermined
longitudinal momentum of the neutrino. The angular distributions have
been calculated in the Collins-Soper frame.  For $W$ decay, this frame
has the unique advantage that the polar and azimuthal angles $\theta$
and $\phi$ can be reconstructed   modulo a sign ambiguity in
$\cos\theta$ without information on the longitudinal momentum of the
neutrino.

When cuts are imposed on the leptons, the only polarization effects
visible in the $\phi$ and $\cos\theta$ distributions occur at large
values of the gauge boson transverse momentum, where unfortunately,
the cross section is smallest.  Polarization effects can be maximized
by minimizing the cuts, however, this strategy is severly limited
since cuts are needed to reject  background events.  Alternatively, it
may be possible to highlight gauge boson polarization effects by
``dividing out'' the kinematic effects, {\it i.e.}, if the
histogrammed data is divided by the theoretical result for isotropic
gauge boson decay, the resulting ratio is more sensitive to
polarization effects.

Since polarization effects are nearly obscured by the  kinematical
effect of cuts,  experimental analyses of the very small so-called
``T-odd'' effects  discussed in Ref.~\cite{npb} or possible $CP$
violation effects in the Standard Model through the Kobayashi-Maskawa
mechanism in hadronic $W$ and $Z$ production \cite{nachtmann1} appear
to be impractical in the presence of realistic cuts at the Tevatron.

\begin{center}
{\bf Acknowledgements}
\end{center}

We thank V.~Barger, F.~Halzen, and D.~Summers for useful discussions.
This work is supported in part by the U.S. Department of Energy under
contract Nos. DE-AC02-76ER00881 and DE-FG03-91ER40674,  by Texas
National Research Laboratory Grant No.~RGFY93-330, and by the
University of Wisconsin Research Committee with funds granted by the
Wisconsin Alumni Research Foundation.

\newpage
\begin{center}
{\bf FIGURE CAPTIONS}
\end{center}
\begin{itemize}
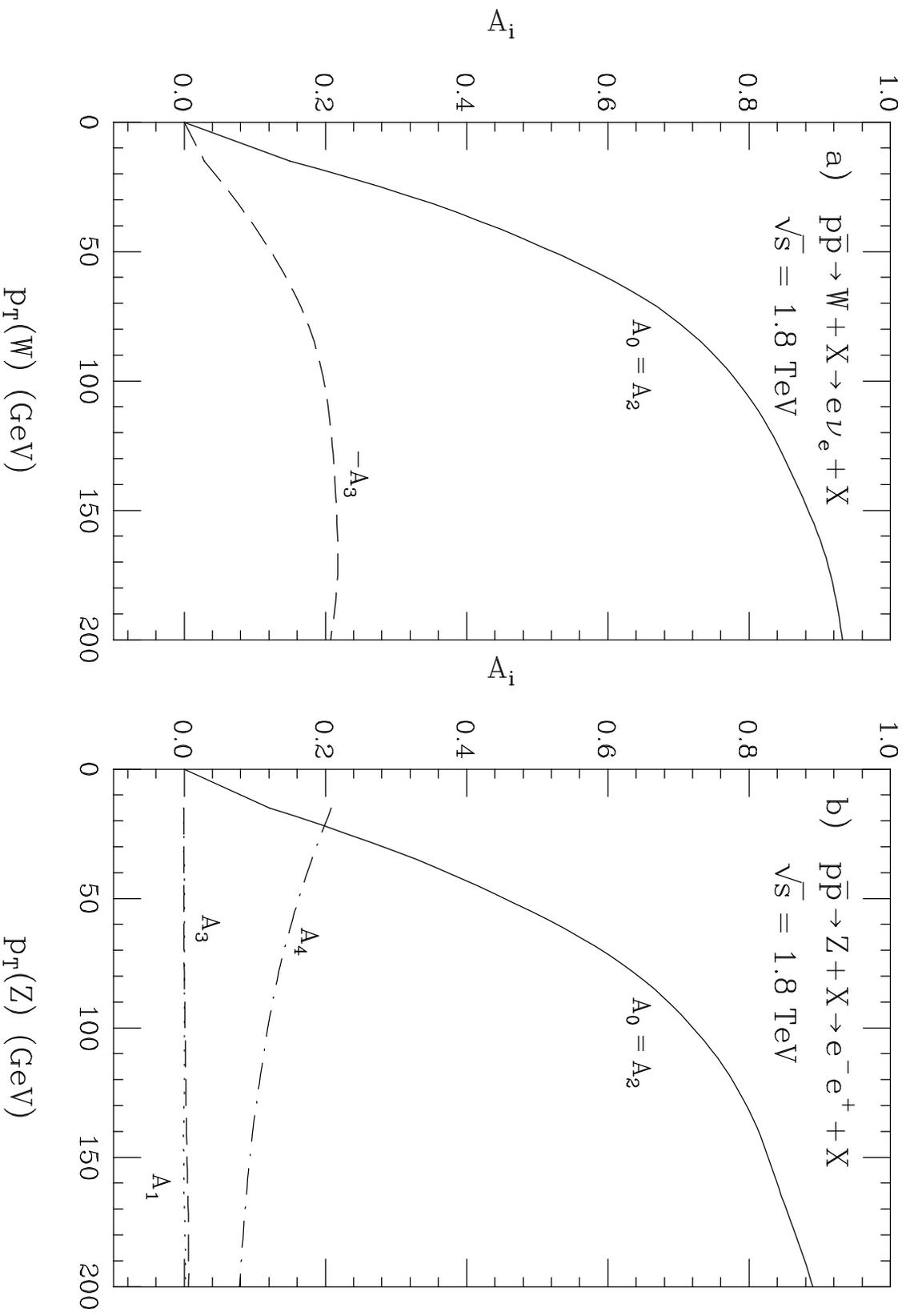
\item[{\bf Fig. 1}]
a) Angular coefficients $A_0, A_2$, and $A_3$ for $W$ boson production
and decay in the CS frame as a function of the transverse momentum
$p_T^{}(W)$ at $\sqrt{S}=1.8$ TeV.\\
b) Angular coefficients $A_0, A_1, A_2, A_3$, and $A_4$ for $Z$ boson
production and decay in the CS frame  as a function of the transverse
momentum $p_T^{}(Z)$ at $\sqrt{S}=1.8$ TeV.\\
No cuts or smearing have been applied.
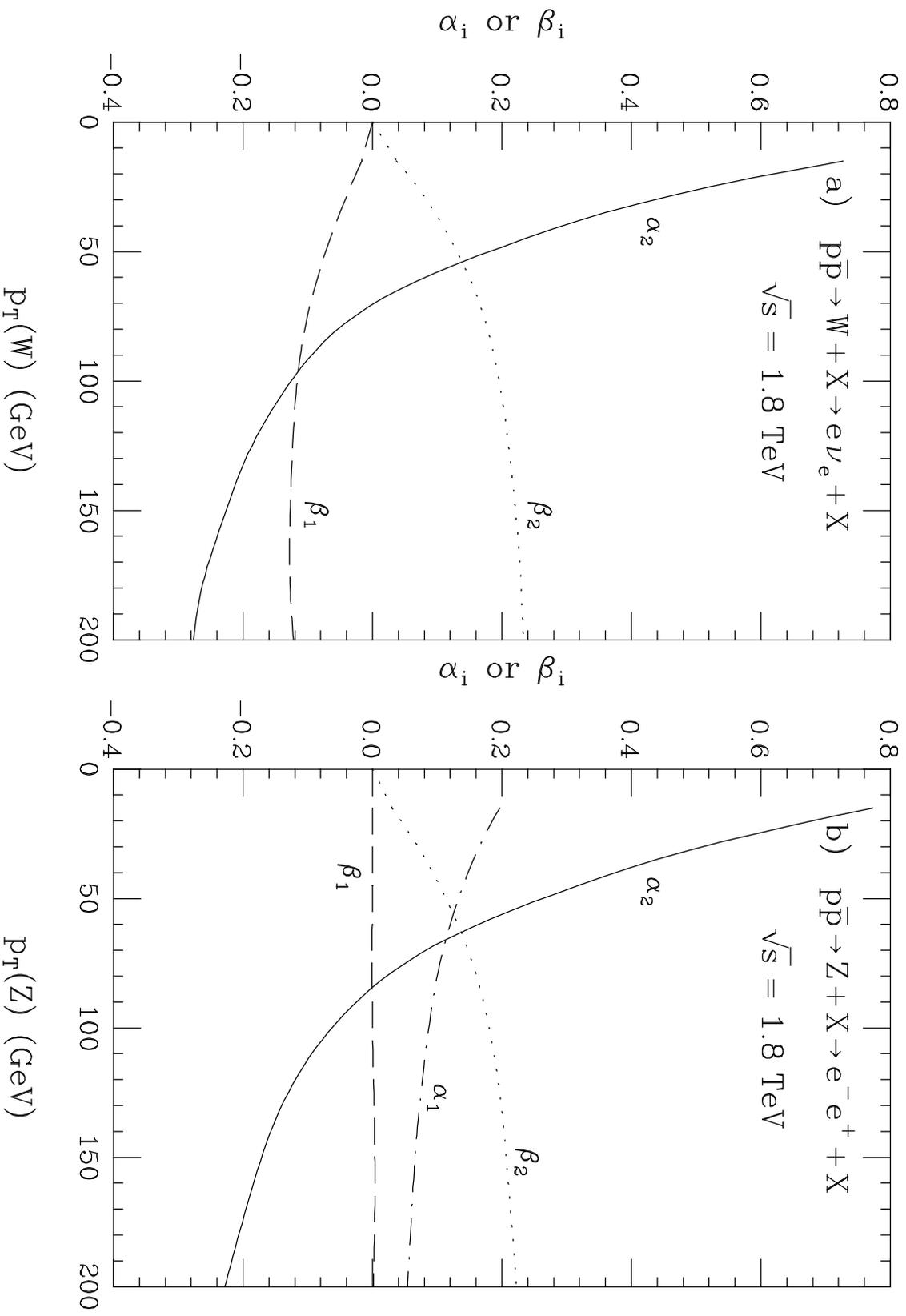
\item[{\bf Fig. 2}]
a) Angular coefficients $\alpha_2, \beta_1$, and $\beta_2$  for $W$
boson production and decay in the CS frame  as a function of the
transverse momentum $p_T^{}(W)$ at $\sqrt{S}=1.8$ TeV.\\
b) Angular coefficients $\alpha_1,\alpha_2,\beta_1$, and $\beta_2$ for
$Z$ boson production and decay in the CS frame  as a function of the
transverse momentum $p_T^{}(Z)$ at $\sqrt{S}=1.8$ TeV.\\
No cuts or smearing have been applied.
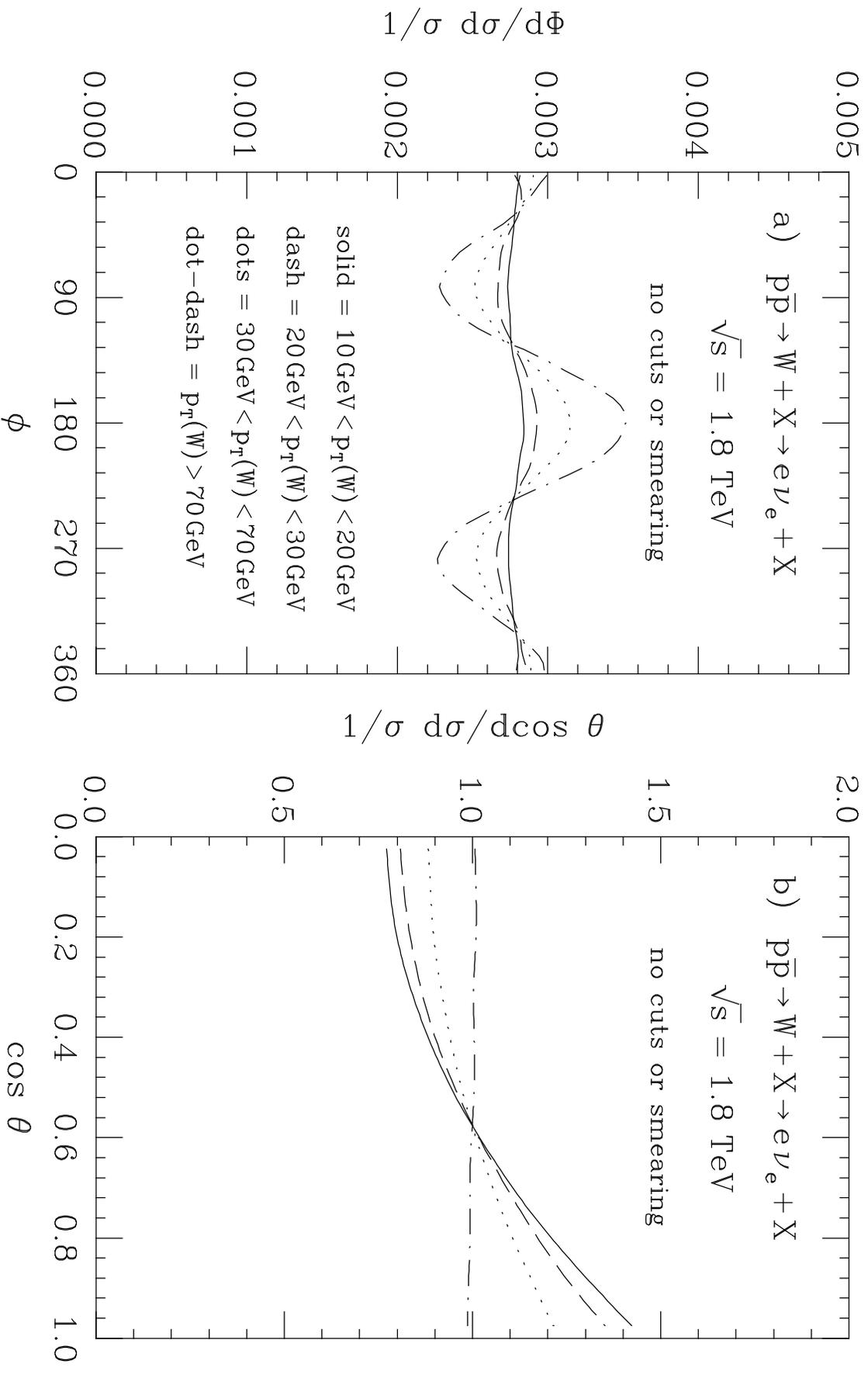
\item[{\bf Fig. 3}]
a) Normalized $\phi$ and b) normalized $\cos\theta$ distributions
of the charged lepton from $W$ boson decay in the CS frame.
Results are shown for four bins in $p_T^{}(W)$:\\
10~GeV $<\, p_T^{}(W)\, < 20$~GeV (solid),\\
20~GeV $<\, p_T^{}(W)\, < 30$~GeV (dashed),\\
30~GeV $<\, p_T^{}(W)\, < 70$~GeV (dots),\\
70~GeV $<\, p_T^{}(W)\, $ (dot-dashed).\\
No cuts or smearing have been applied.
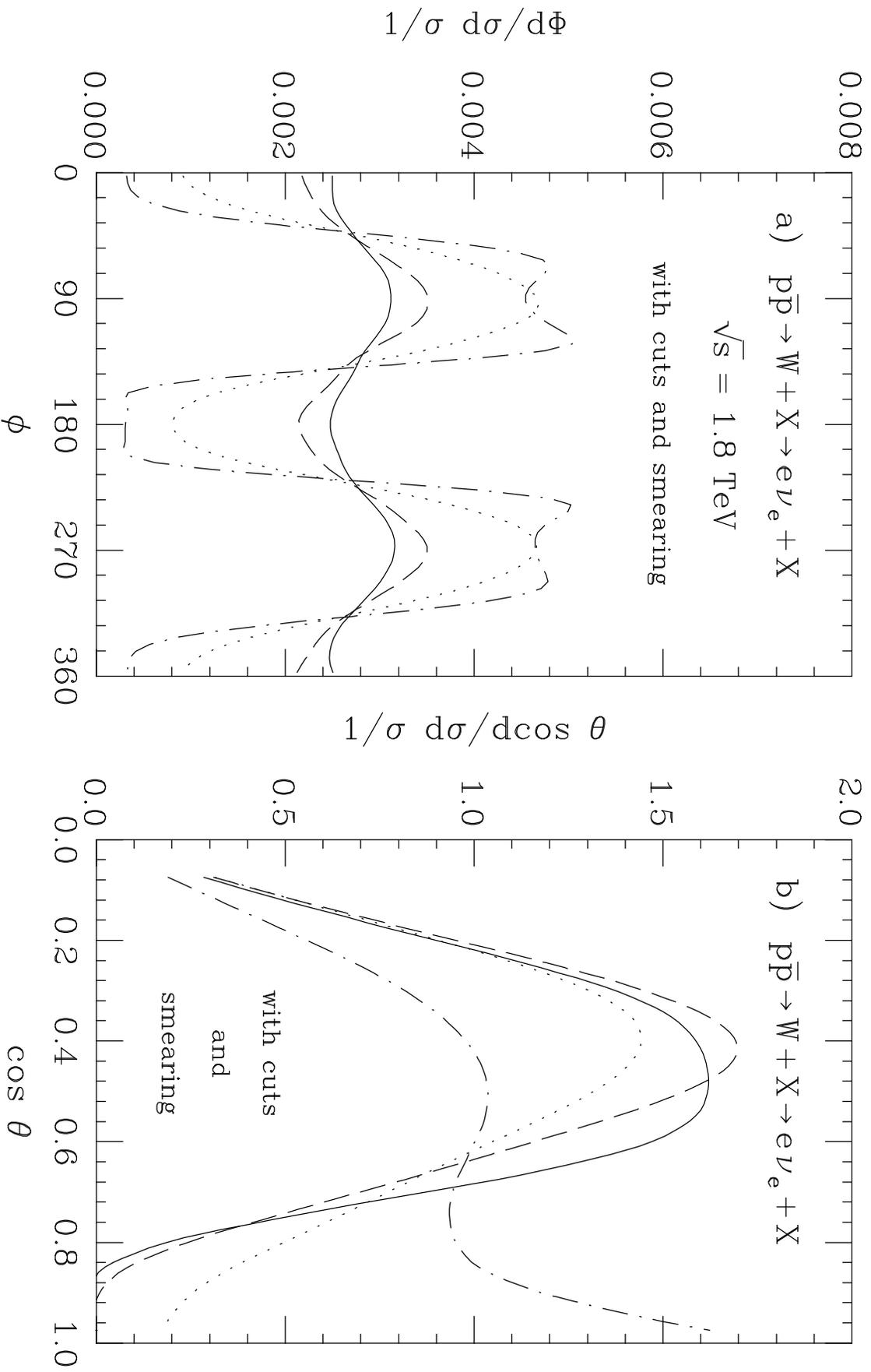
\item[{\bf Fig. 4}]
Same as Fig.~3 but with smearing and the cuts
$p_T^{}(l) > 25$~GeV, $p\llap/_T^{} > 25$~GeV, and $y(l) < 1$.
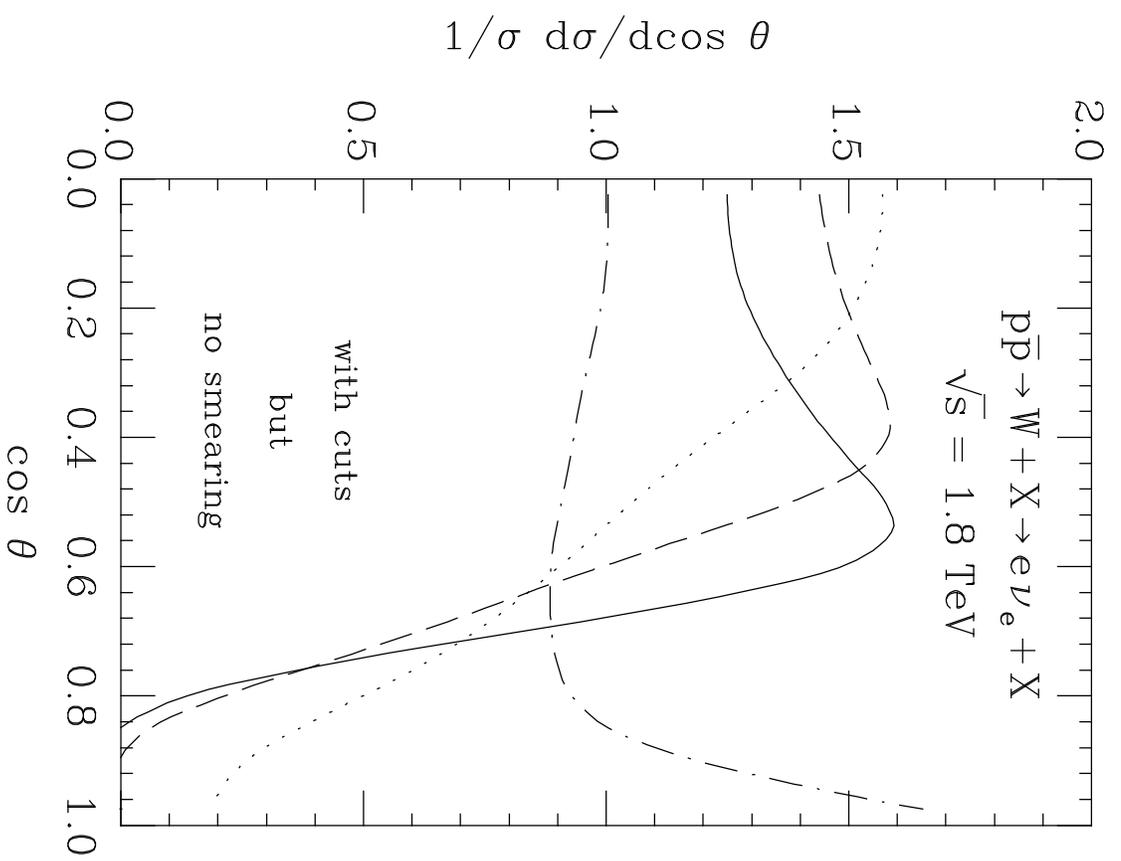
\item[{\bf Fig. 5}]
Normalized $\cos\theta$ distribution of the charged lepton from $W$
boson decay in the CS frame. The cuts $p_T^{}(l) > 25$~GeV,
$p\llap/_T^{} > 25$~GeV, and $y(l) < 1$ have been imposed, but no
smearing is included.
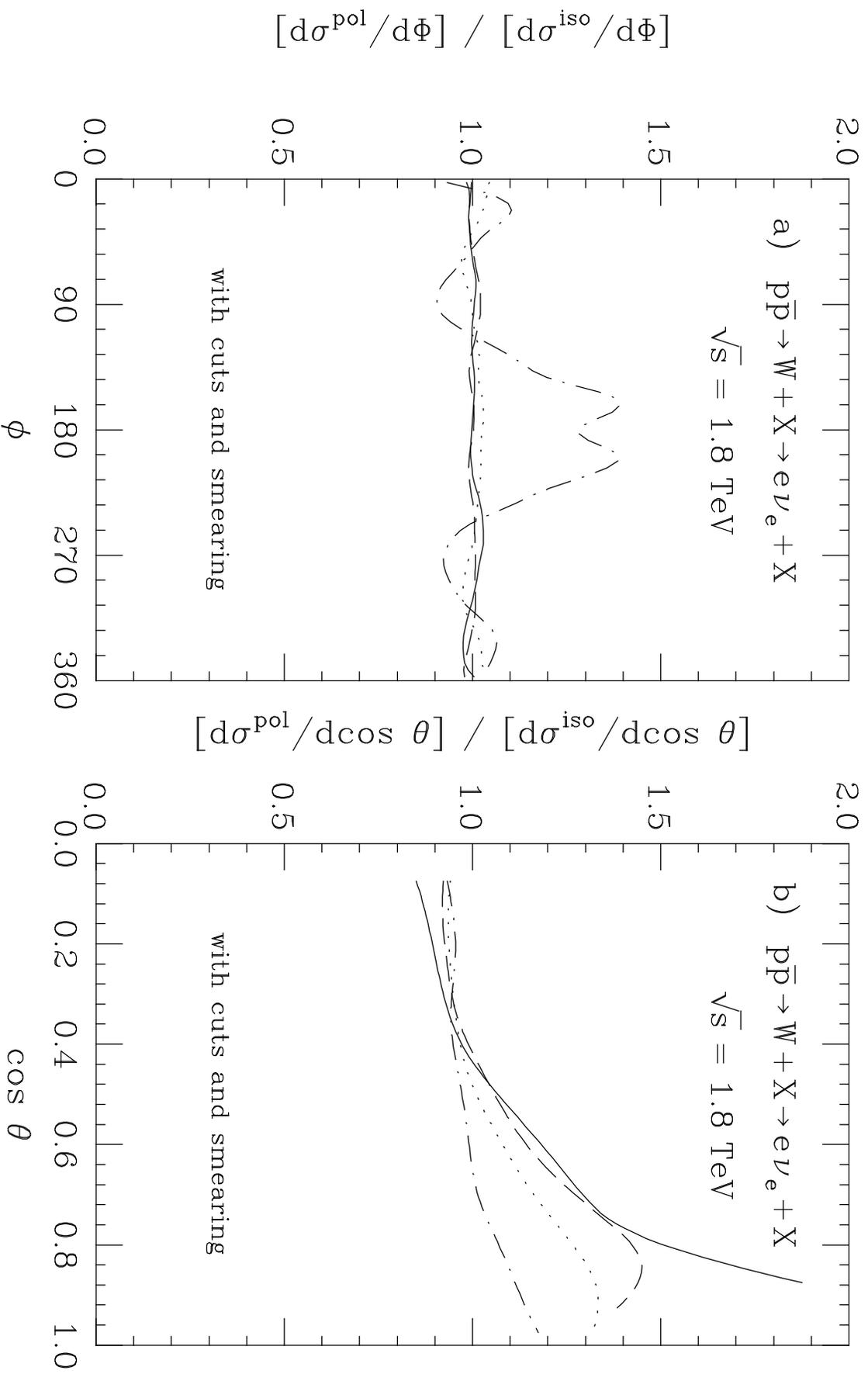
\item[{\bf Fig. 6}]
Ratios of distributions obtained with full polarization effects to
those obtained with isotropic decay of the $W$ boson. Parts a) and b)
are the ratios for the $\phi$ and $\cos\theta$ distributions,
respectively. Energy resolution smearing and the cuts $p_T^{}(l) >
25$~GeV, $p\llap/_T^{} > 25$~GeV, and $y(l) < 1$ are included.
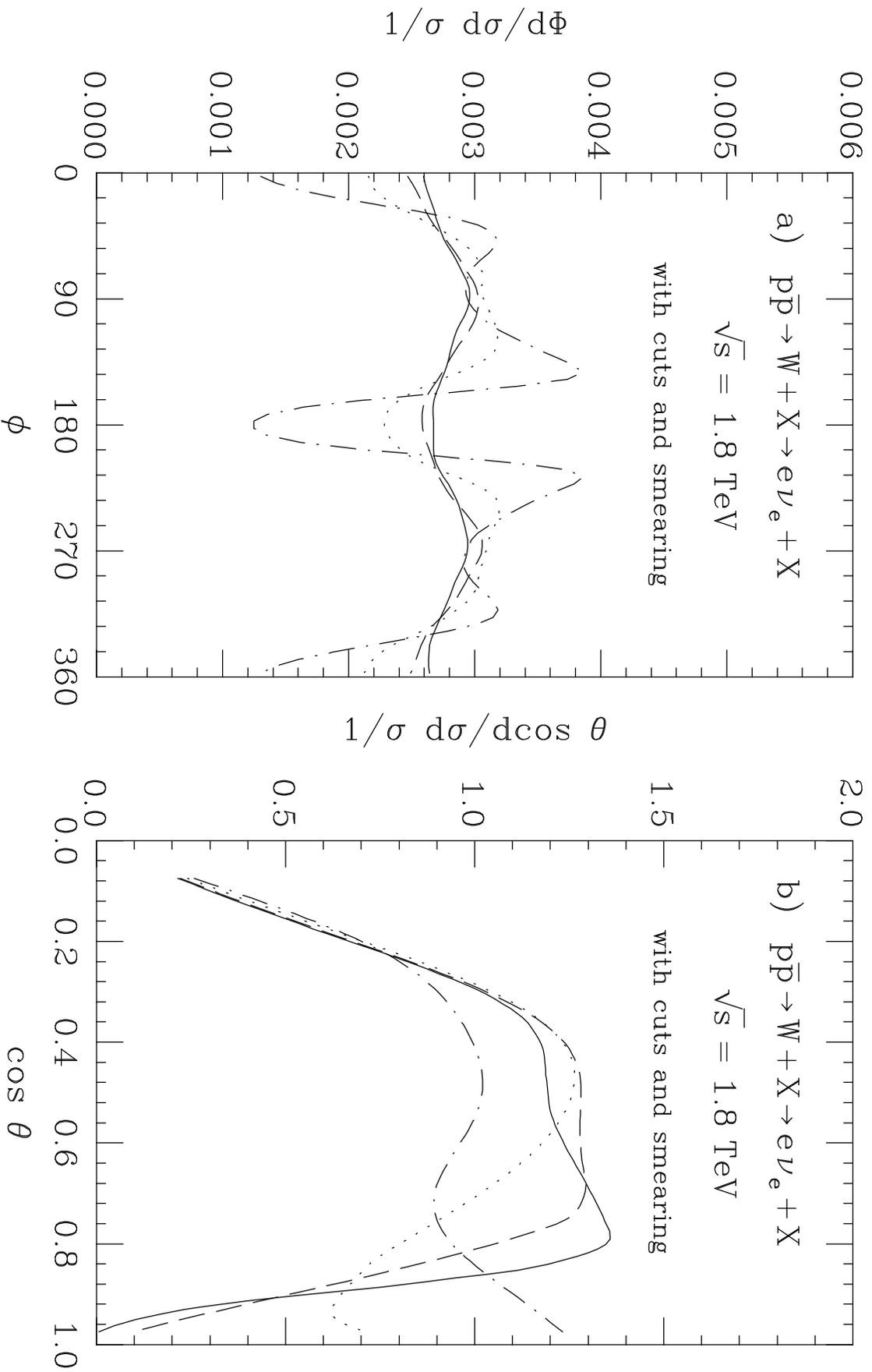
\item[{\bf Fig. 7}]
Same as Fig.~3 but with smearing and the cuts
$p_T^{}(l) > 15$~GeV, $p\llap/_T^{} > 15$~GeV, and $y(l) < 2.5$.
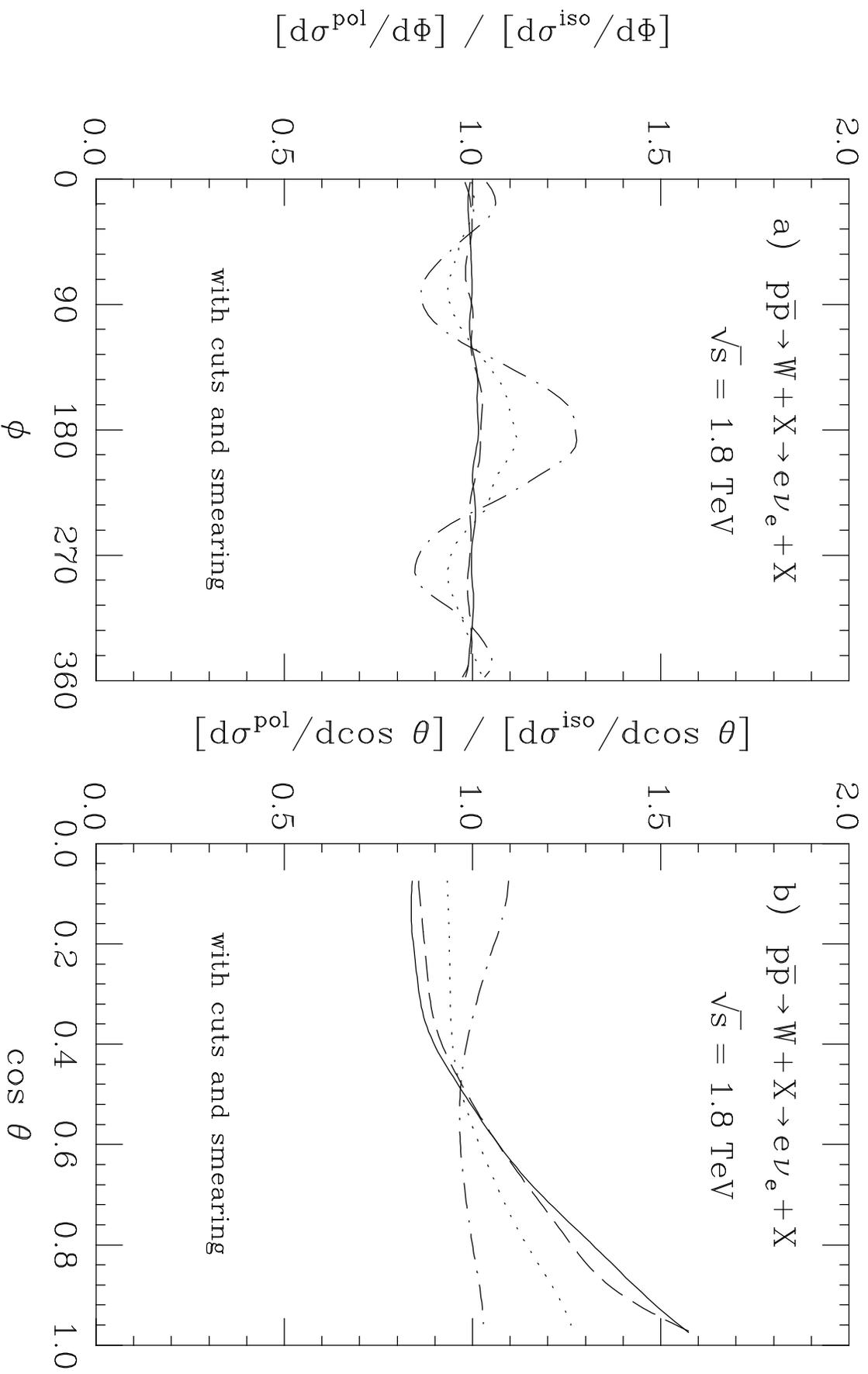
\item[{\bf Fig. 8}]
Same as Fig.~6 but with the cuts
$p_T^{}(l) > 15$~GeV, $p\llap/_T^{} > 15$~GeV, and $y(l) < 2.5$.
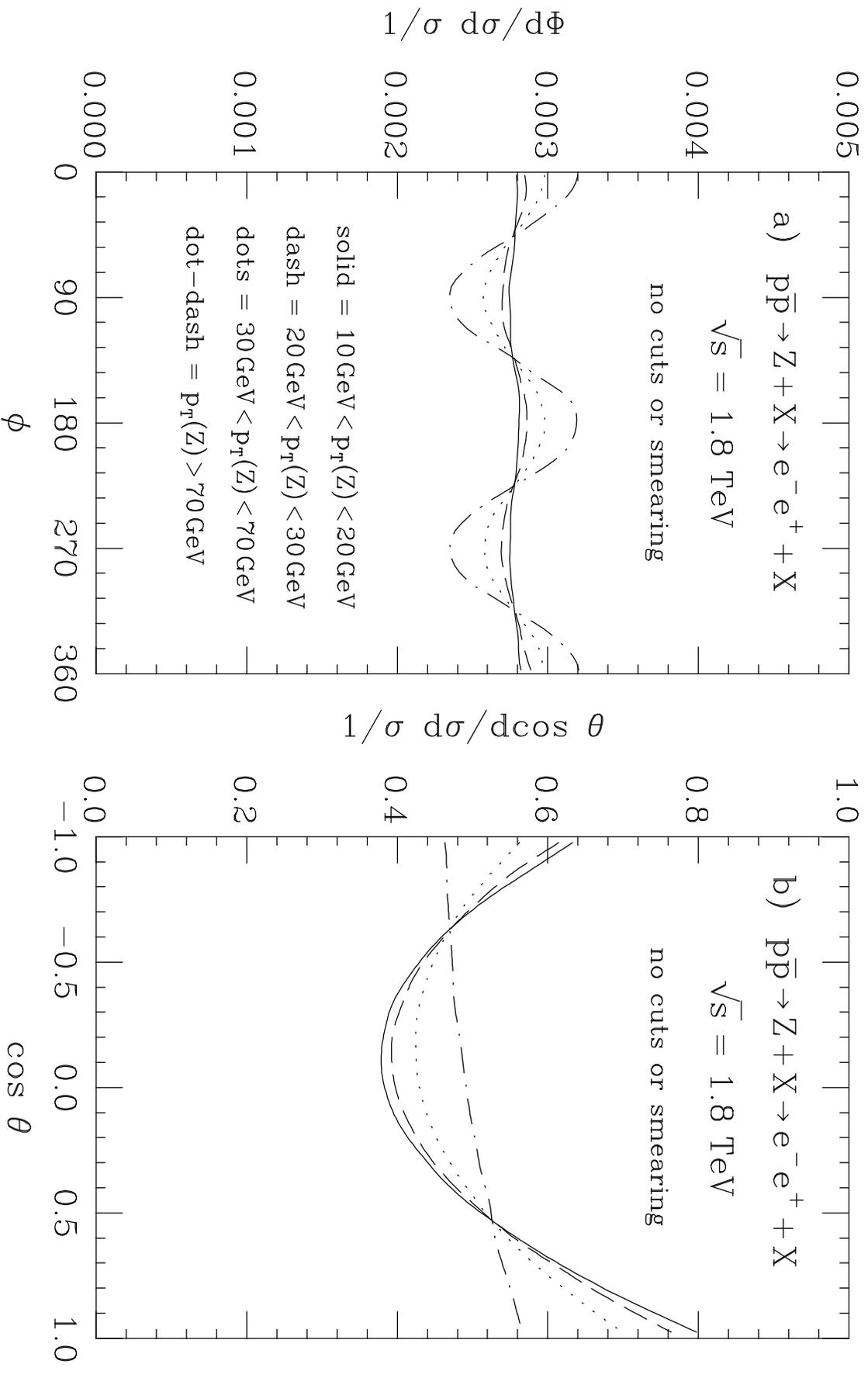
\item[{\bf Fig. 9}]
a) Normalized $\phi$ and b) normalized $\cos\theta$ distributions  of
the negatively charged lepton from $Z$ boson decay in the CS frame.
Results are shown for four bins in $p_T^{}(Z)$:\\
10~GeV $<\, p_T^{}(Z)\, < 20$~GeV (solid),\\
20~GeV $<\, p_T^{}(Z)\, < 30$~GeV (dashed),\\
30~GeV $<\, p_T^{}(Z)\, < 70$~GeV (dots),\\
70~GeV $<\, p_T^{}(Z)\, $ (dot-dashed).\\
No cuts or smearing have been applied.
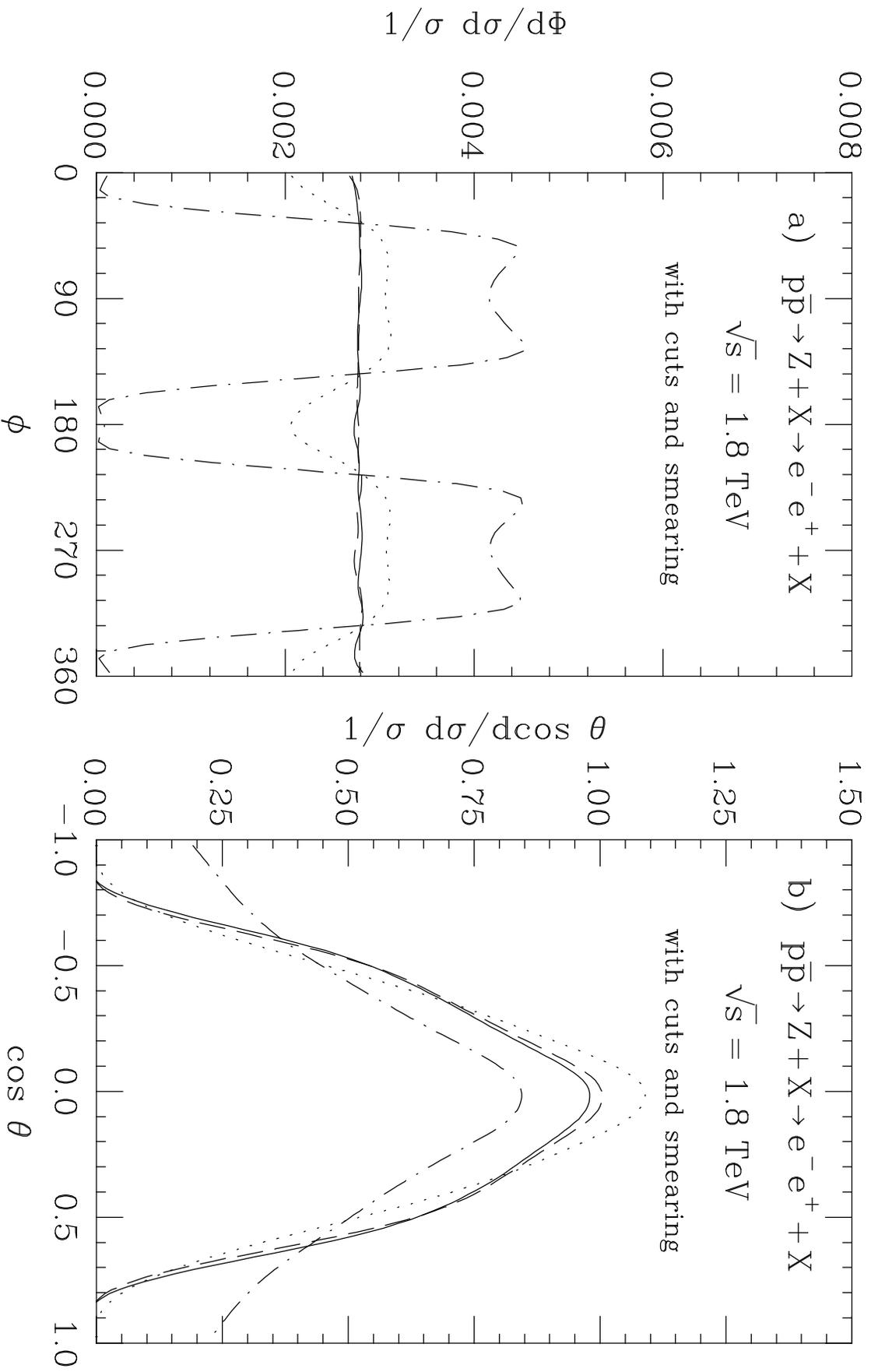
\item[{\bf Fig. 10}]
Same as Fig.~9 but with smearing and the cuts
$p_T^{}(l) > 25$~GeV and $y(l) < 1$.
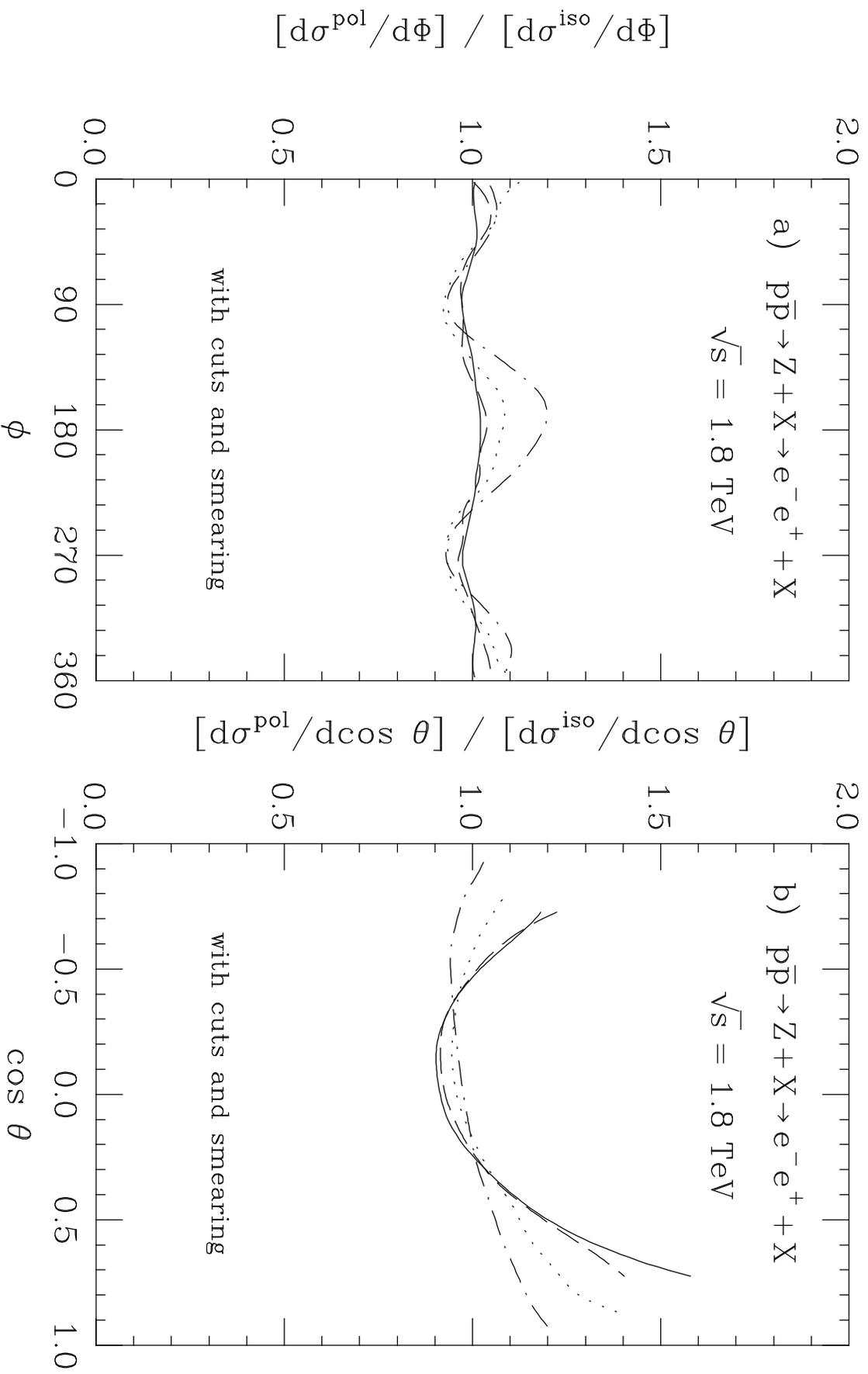
\item[{\bf Fig. 11}]
Ratios of distributions obtained with full polarization effects to
those obtained with isotropic decay of the $Z$ boson. Parts a) and b)
are the ratios for the $\phi$ and $\cos\theta$ distributions,
respectively. Energy resolution smearing and the cuts $p_T^{}(l) >
25$~GeV and $y(l) < 1$ are included.
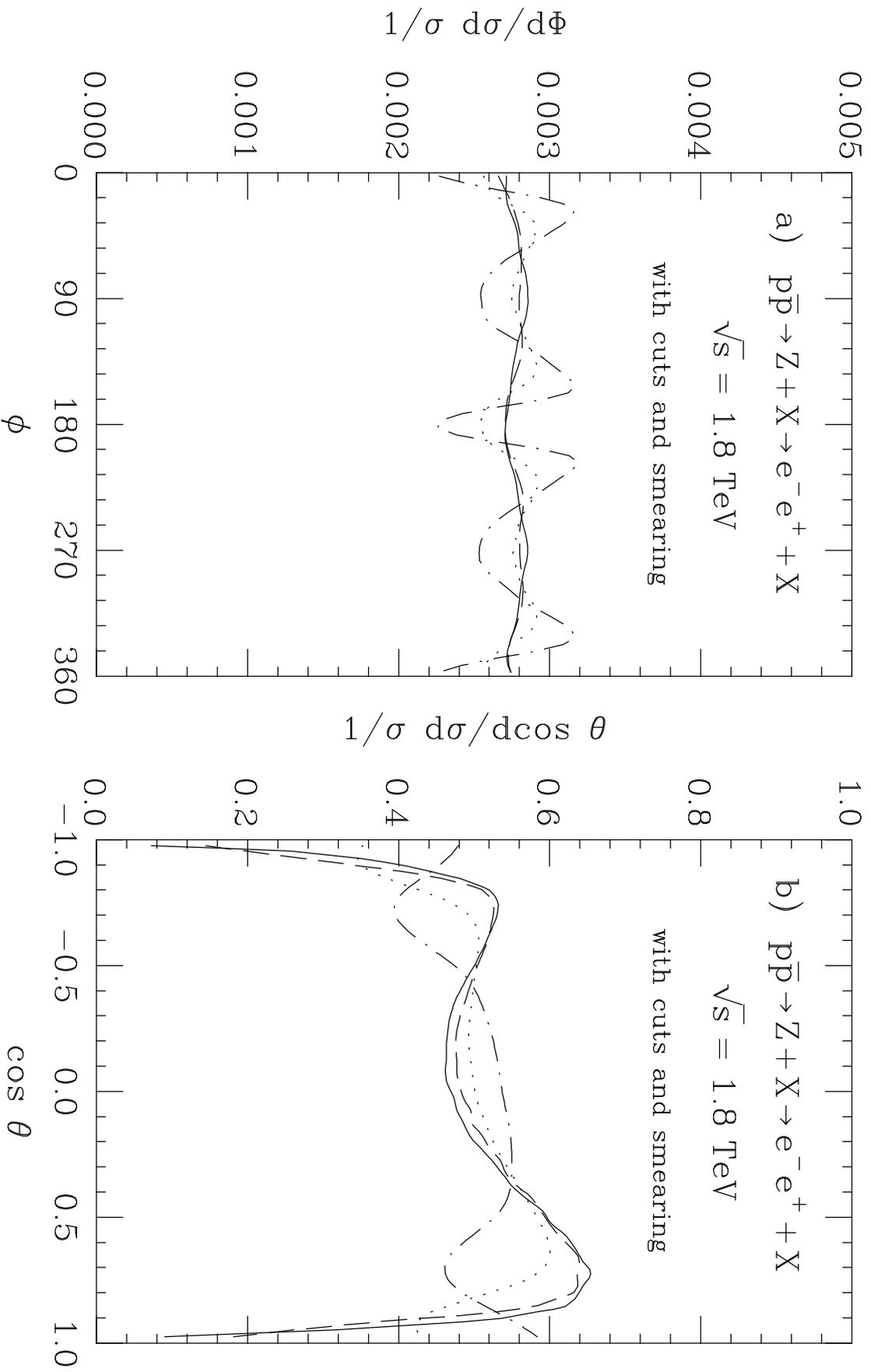
\item[{\bf Fig. 12}]
Same as Fig.~9 but with smearing and the cuts
$p_T^{}(l) > 10$~GeV and $y(l) < 2.5$.
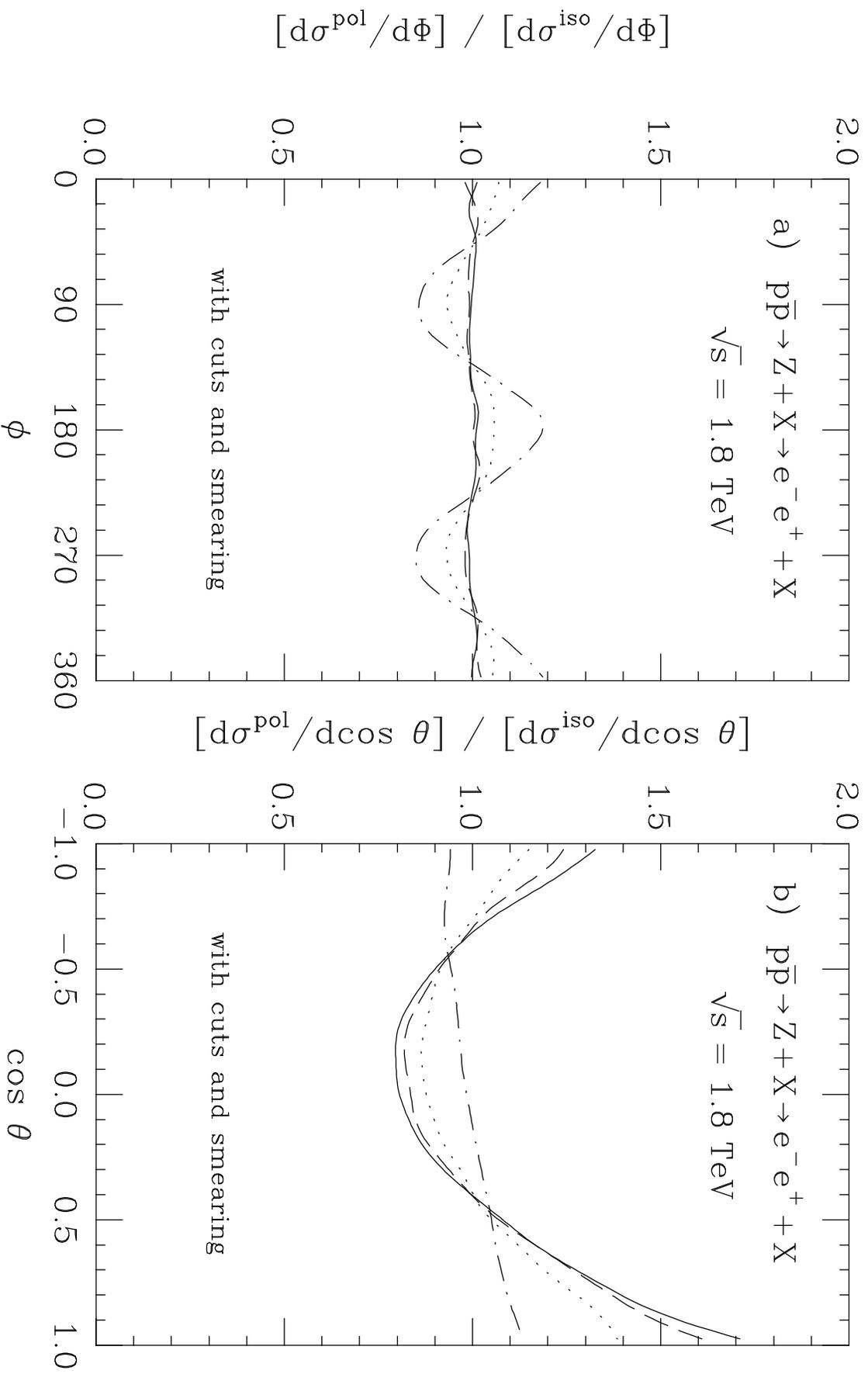
\item[{\bf Fig. 13}]
Same as Fig.~11 but with the cuts
$p_T^{}(l) > 10$~GeV and $y(l) < 2.5$.
\end{itemize}

\def\npb#1#2#3{{\it Nucl. Phys. }{\bf B #1} (#2) #3}
\def\plb#1#2#3{{\it Phys. Lett. }{\bf B #1} (#2) #3}
\def\prd#1#2#3{{\it Phys. Rev. }{\bf D #1} (#2) #3}
\def\prl#1#2#3{{\it Phys. Rev. Lett. }{\bf #1} (#2) #3}
\def\prc#1#2#3{{\it Phys. Reports }{\bf C #1} (#2) #3}
\def\pr#1#2#3{{\it Phys. Reports }{\bf #1} (#2) #3}
\def\zpc#1#2#3{{\it Z. Phys. }{\bf C #1} (#2) #3}
\def\ptp#1#2#3{{\it Prog.~Theor.~Phys.~}{\bf #1} (#2) #3}
\def\nca#1#2#3{{\it Nouvo~Cim.~}{\bf A #1} (#2) #3}
\newpage
\sloppy
\raggedright

\end{document}